\documentclass[final,3p,10pt,sort&compress]{elsarticle}
\usepackage[english]{babel}

\usepackage{graphicx}
\usepackage{natbib}
\usepackage{rotating}
\usepackage{subfigure}

\usepackage{amsmath}
\usepackage{amsfonts}
\usepackage{amssymb}

\usepackage{url}
\usepackage{multirow}
\usepackage{acronym}
\usepackage[algoruled]{algorithm2e}

\usepackage{color}

\usepackage{ifpdf}

\usepackage{colortbl}
\usepackage{lscape}
\usepackage{pdflscape} 
\usepackage{array,booktabs,ragged2e}

\usepackage{xcolor}
\definecolor{lightgrey}{RGB}{219, 219, 219}
\definecolor{verylightblue}{RGB}{204, 229, 255}
\definecolor{lightblue}{RGB}{124, 216, 255}
\definecolor{blue}{RGB}{32, 187, 253}

\newcolumntype{R}[1]{>{\RaggedLeft\arraybackslash}p{#1}}

\journal{ArXiv}

\begin{document}

\begin{frontmatter}

\title{An InfoVis Tool for Interactive Component-Based Evaluation}

\author{Giacomo Rocco}

\author{Gianmaria Silvello}
\ead{silvello@dei.unipd.it}

\address{Department of Information Engineering, University of Padua, Italy.}

\begin{abstract}
In this paper, we present an InfoVis tool based on Sankey diagrams for the exploration of large combinatorial combinations of IR components -- the \acf{GoP}. The goal of this tool is to ease the comprehension of the behavior of single IR components within fully functioning off-the-shelf IR systems without recurring to complex statistical tools.
\end{abstract}


\end{frontmatter}

\section{Motivations}
\ac{IR} systems are constituted of ``pipelines'' of components such as stop lists, stemmers and \ac{IR} models, which are stacked together in order to process both documents and user queries and to match them returning a ranked result list of documents in decreasing order of estimated relevance. The performance of \ac{IR} systems are evaluated in terms of \emph{effectiveness} that can be determined only after that the system has been built; indeed, no effectiveness prediction about a specific component can be done before it has been tested within a fully functioning IR system. 

Currently, the only viable means to determine the contribution to the system effectiveness of single components is to measure their impact on the overall performances by testing all the different combinations of such components. This leads to a very high number of cases to be considered, making the space of system combinations large and complex to explore.

Besides requiring a great deal of effort and resources to be produced, these combinatorial compositions constitute a challenge when it comes to explore, analyze, and make sense of the experimental results with the goal of understanding how different components contribute to the overall performances and interact together. Indeed, it is typically needed to resort to rather complex statistical tools (e.g. multi-way \ac{ANOVA} models) requiring a careful experimental design and producing results which call for a considerable extent of expertise to be interpreted~\cite{FerroSilvello2016}. To this end, we developed an extensive set of $612\times6 = 3,672$ systems -- i.e. the \acf{GoP} \footnote{\url{http://gridofpoints.dei.unipd.it/}} -- arising from the combinatorial composition of several open-source publicly available components such as stop lists, stemmers, and \ac{IR} models, and run against $6$ different public test collections shared by the \ac{TREC} international evaluation initiative. Thanks to this \ac{GoP}, in~\cite{FerroSilvello2017b} we presented the deep statistical analyses we run and the insights we gathered about the individual contributions of single IR components to the overall performances of fully working IR systems.

In this paper we present an InfoVis system based on SanKey diagrams -- often used in physics to represent energy inputs, useful output, and wasted output -- to allow the exploration of the \ac{GoP} to quickly understand which combinations perform best under specific criteria, how components behave across a wide range of cases, and how they interact together. Our main goal is to give \ac{IR} researchers and practitioners a fast and easy way to understand and analyze the \ac{GoP} without recurring to demanding and complex statistical tools. 

Hence, the InfoVis tool we present enables the analysis and comparison of a complex set of measures associated with a large combinatorial space of \ac{IR} systems and the intuitive exploration and understanding of many component configurations. It is thought to be  simple to use and to favor interaction, thus it provides functionalities as component filtering, measure selection and tooltips presenting statistical information easy to interpret. 

\section{Related Work}
InfoVis techniques are typically exploited for the presentation and exploration of the {\it documents} managed by an \ac{IR} system~\cite{Zhang2008}. Typical examples are: identification of the objects and their attributes to be displayed~\cite{FowlerEtAl1991}; different ways of presenting the data~\cite{MorseEtAl2002}; the definition of visual spaces and visual semantic frameworks~\cite{Zhang2001}. The development of interactive means for \ac{IR} is an active field which focuses on search user interfaces~\cite{Hearst2011}, displaying of results~\cite{CrestaniEtAl2004} and browsing capabilities~\cite{Koshman2005}.

Less attention has been dedicated to the application of InfoVis techniques to the analysis of experimental evaluation results. One example of a system applying visualization to \ac{IR} is \ac{VIRTUE}, a visual analytics tool supporting performance and failure analysis~\cite{AngeliniEtAl2014}. In the same vein,~\cite{AngeliniEtAl2016} presents an analytical framework trying to learn the behavior of a system just from its outputs for obtaining a rough estimation of the possible effects of a modification to the system. More recently,~\cite{LipaniEtAl2017} presented an InfoVis tool to explore pooling strategies.

\begin{figure*}[th!]
\centering
\includegraphics[width=1\textwidth]{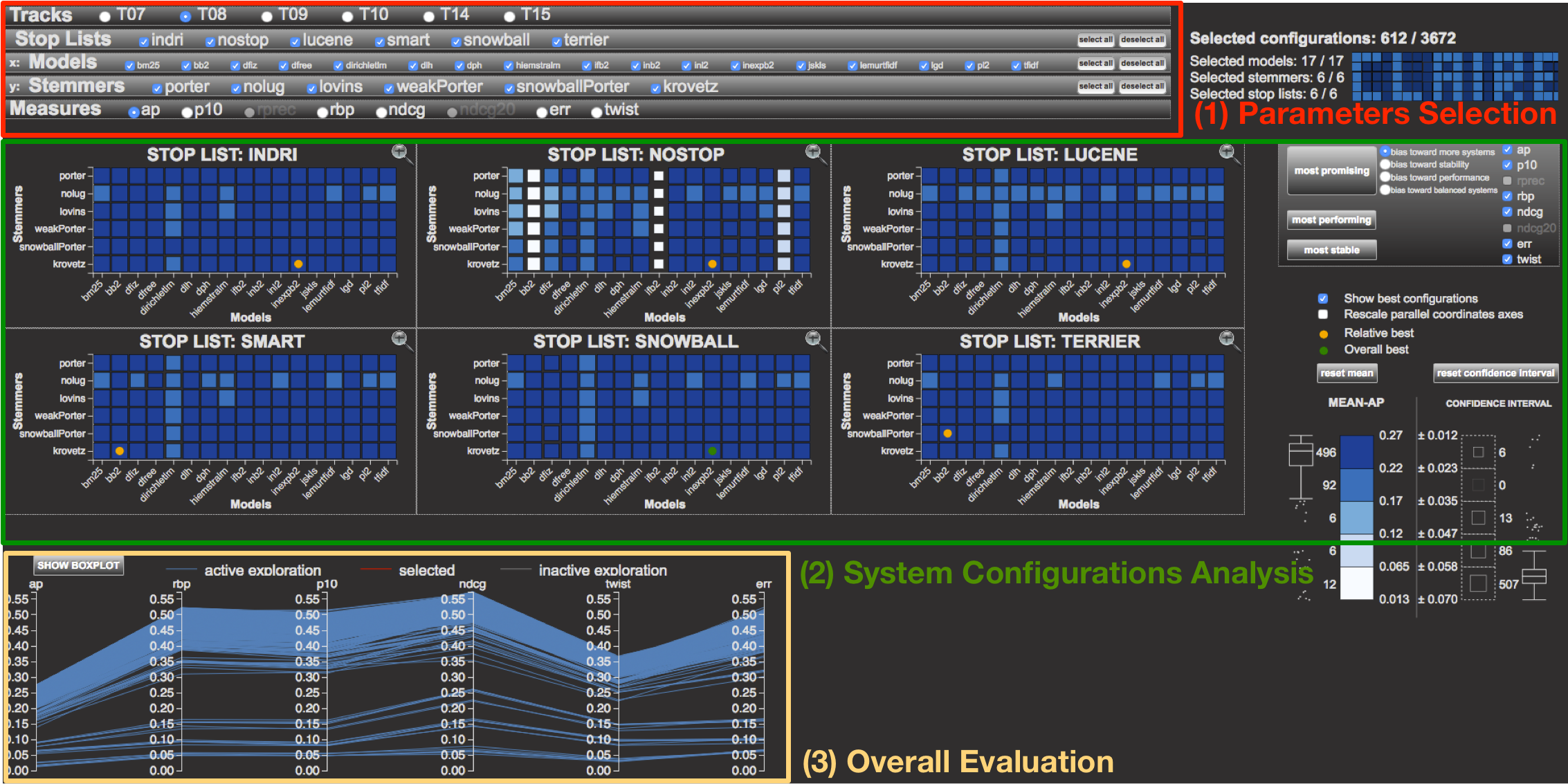}
\caption{The overall view of the CLAIRE visual analytics tool~\cite{AngeliniFFSS18}. \label{fig:CLAIRE}}
\end{figure*}

However, to the best of our knowledge only one solution -- i.e. the CLAIRE tool~\cite{AngeliniFFSS18}, see Figure~\ref{fig:CLAIRE} -- exists for dealing with large sets of \ac{IR} systems -- the \ac{GoP}~\cite{FerroSilvello2016e,FerroSilvello2016} -- generated by many \ac{IR} components which allows the inspection of both configurations and measures. CLAIRE is based on a totally different visual paradigm since it uses tiles, parallel coordinates and boxplots to explore system configurations.

\section{Experimental Setting}

The \ac{GoP} data adopted by our InfoVis tool is based on three main components of an IR system: stop list, stemmer, and \ac{IR} model. We selected a set of alternative implementations of each component and, by using the Terrier v.4.0 \footnote{\url{http://www.terrier.org/}} open source system, we created a run for each system defined by combining the available components in all possible ways. The selected components are: 
\begin{itemize}
	\item \emph{Stop list}:~\texttt{nostop}, \texttt{indri}, \texttt{lucene}, \texttt{snowball}, \texttt{smart},\texttt{terrier};
	\item \emph{Stemmer}: \texttt{nolug}, \texttt{weakPorter}, \texttt{porter},
	 \texttt{snowballPorter}, \\ \texttt{krovetz}, \texttt{lovins};
	\item \emph{Model}:  \texttt{bb2}, \texttt{bm25}, \texttt{dfiz}, \texttt{dfree}, \texttt{dirichletlm}, \texttt{dlh}, \texttt{dph}, \\ \texttt{hiemstralm}, \texttt{ifb2}, \texttt{inb2}, \texttt{inl2}, \texttt{inexpb2}, \texttt{jskls}, \texttt{lemurtfidf}, \texttt{lgd}, \texttt{pl2}, \texttt{tfidf}.
\end{itemize}

Overall, these components define a $6 \times 6 \times 17 = 612$ runs.  The stop lists differ from each other by the number of terms composing them; specifically, \texttt{indri} has 418 terms, \texttt{lucene} has 33 terms, \texttt{snowball} has 174 terms, \texttt{smart} has 571 terms and \texttt{terrier} 733 terms. Stemmers can be classified into aggressive (e.g. \texttt{lovins}) and weaker stemmers (e.g. \texttt{porter}).

The models we employ are classified into the three main approaches currently adopted by search engines: (1) the vector space model -- e.g. \texttt{tfidf} and \texttt{lemurtfidf}; (2) the probabilistic model -- e.g. \texttt{bm25} and the \ac{DFR} models; and, (3) the language models -- e.g. \texttt{dirichletlm}, \texttt{hiemstralm} and \texttt{lgd}. 
We considered $6$ standard and shared collections with 50 different topics each: \emph{TREC Adhoc tracks \emph{\texttt{T07}} and \emph{\texttt{T08}}}; \emph{TREC Web tracks  \emph{\texttt{T09}} and \emph{\texttt{T10}}}; and, \emph{TREC Terabyte tracks  \emph{\texttt{T14}} and \emph{\texttt{T15}}}.
We evaluate the \acp{GoP} by employing $8$ evaluation measures: \acs{AP}, P@10, Rprec, \acs{RBP}, \acs{nDCG}, nDCG@20,  \acs{ERR}, and Twist. 

Summarizing, the \ac{GoP} we visualize with the proposed InfoVis tool consists of 612 runs over 6 collections with 50 topics each and evaluated with 8 measures, which amounts to almost 1.5M data points.

\section{The InfoVis tool}
\begin{figure*}[th!]
\centering
\includegraphics[width=1\textwidth]{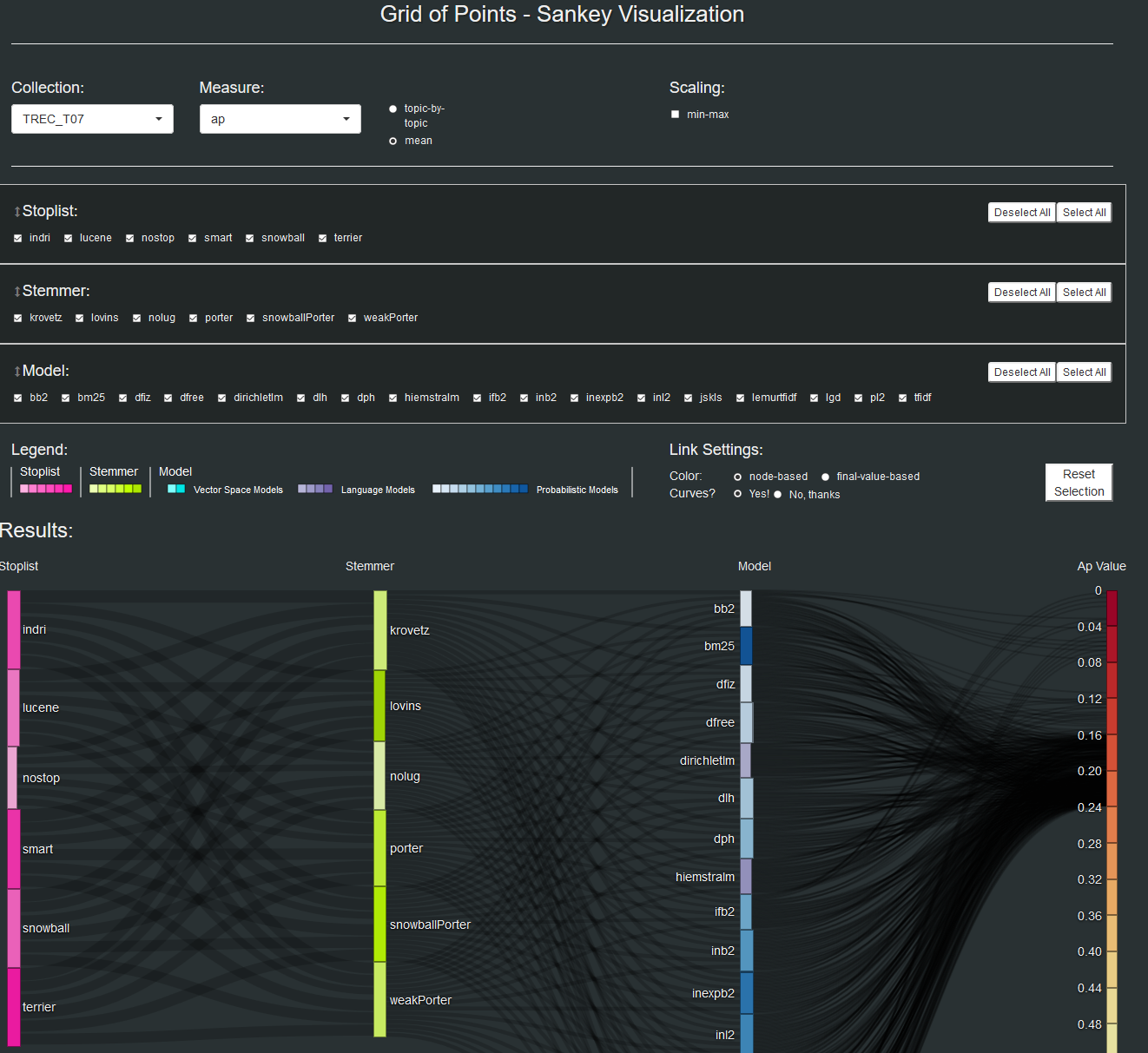}
\caption{The overall InfoVis system; on the top there is the parameter selection area and on the bottom the dynamic SanKey diagram. \label{fig:SANKEY}}
\end{figure*}

The InfoVis tool we realized, see Figure \ref{fig:SANKEY} is composed of two main areas: 
\begin{description}
\item[Parameters selection area:](top of Figure \ref{fig:SANKEY}) it allows the user to load the runs relative to the desired experimental collection, to select the components s/he wants to consider and the evaluation measure to be used. 
\item[System analysis area:] (bottom of Figure \ref{fig:SANKEY}) it allows the actual analysis and exploration of the various components and their evaluation on the basis of the parameters selected above.
\end{description}

\subsection{Parameter selection area}
\begin{figure*}[th!]
\centering
\includegraphics[width=1\textwidth]{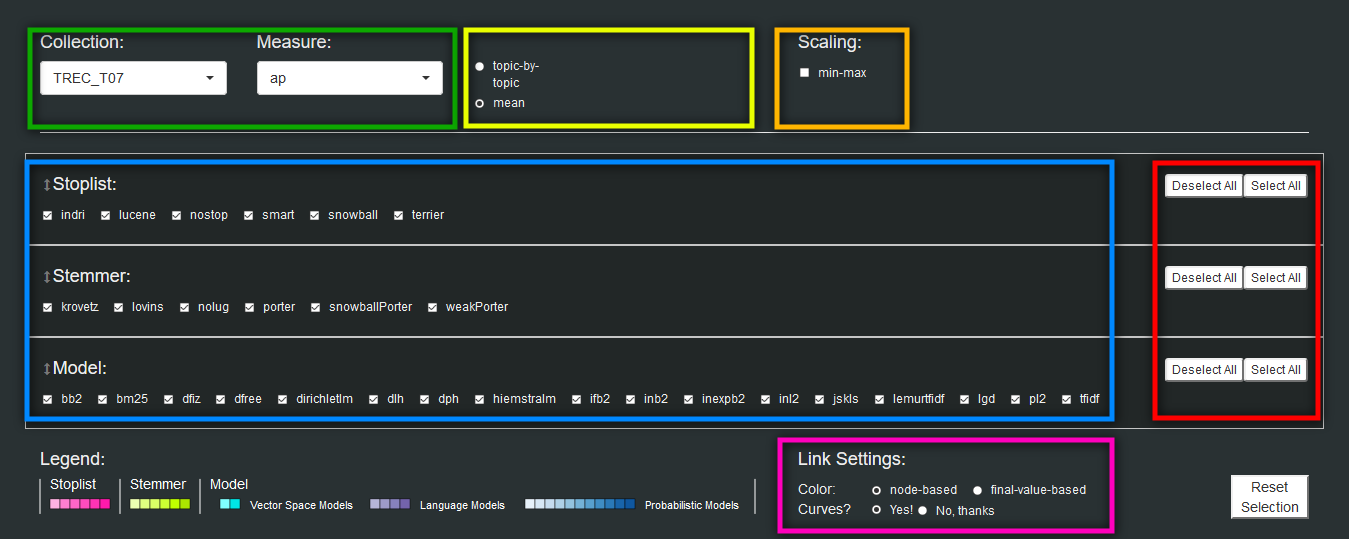}
\caption{A detailed view of the {\emph parameter selection area}. \label{fig:riferimenti}}
\end{figure*}

In Figure \ref{fig:riferimenti} we can see a detailed view of the parameter selection area. 

The first two parameters that can be selected (in the green box) are the experimental collection and the evaluation measure of interest. On the left of these two drop down menus we can choose to visualize the system performances topic-by-topic (if this option is selected a new drop-down menu appears allowing the user to select the topic of interest) or on average (e.g. MAP). The ``scaling'' option enables a normalized visualization of the SanKey diagram (only actual min-max values or the whole range such as $[0,1]$ for AP). The blue box in Figure \ref{fig:riferimenti} shows the control panel enabling the dynamic selection of component families to be visualized in the Sankey diagram. 

The three component families (stoplists, stemmers and IR models) can be re-ordered by a simple drag-and-drop action, leading to a dynamic re-ordering of the axes of the Sankey diagram; this is particularly useful when during the data analysis phase we want to highlight the components interaction. The default axes order better shows the interaction between stoplists and stemmers and between stemmers and IR models, but by re-ordering the axes we can highlight, for instance, the stoplists-models interaction. 

Below the blue box we can see the legend of the Sankey diagram where three chromatic variations are used to differentiate between the components of each family and sub-family of components: fuchsia for stoplists, green for stemmers, light blue for vector space models, purple for language models and dark blue for probabilistic models. The fuchsia box highlights the link settings where we can choose the shape of the SanKey curves and their color schema -- i.e. based on component selection or based on evaluation measure value selection.

Every single interaction with the parameter selection area produces an effect on the SanKey diagram which is rendered dynamically and in real-time; this is intended to ease the interaction with the system and the data analyses to be performed.

\subsection{System analysis area}
\begin{figure*}[th!]
\centering
\includegraphics[width=1\textwidth]{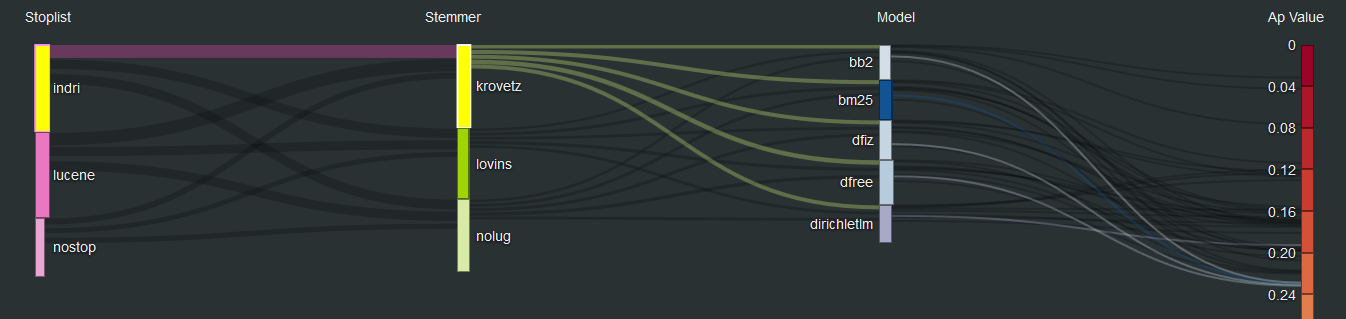}
\caption{A detailed view of the {\emph system analysis area} where some components have been filtered out 
and some other are highlighted for an in-depth analysis of the interactions. \label{fig:analysis}}

\end{figure*}

On the bottom of Figure \ref{fig:SANKEY} we can see the entire analysis space where all the available components are displayed by the SanKey diagram, whereas in Figure \ref{fig:analysis} we can see a restricted analysis area where only some specific components have been selected and highlighted for an in-depth analysis of their performances and interactions. 

The rightmost column presents the evaluation measure values divided into 25 rectangles of equal size, each one representing a $0.04$ value interval. The color of each rectangle follows the red-yellow-green schema where reddish rectangles are assigned to lower values and the greenish ones to higher values. By the means of a drag-and-drop mouse action it is also possible to re-order the rectangles representing family components. Each single link insisting on these rectangles represents one of the 612 systems and their overall performance values. 

A single system is represented by a path, i.e. a series of links connecting one component with the next one. The user can select a set of components (left click on one or more rectangles) to highlight the paths of interest as shown in Figure \ref{fig:analysis} where we selected the \texttt{indri} stoplist and the \texttt{krovetz} stemmer.

The component columns present a number of rectangles equal to the components selected in the {\emph parameter selection area} and the size of the rectangle gives a visual idea of the performances of the component it represents. This is done by calculating the marginal arithmetic mean of the performance values obtained by the systems using a specific component; the means are dynamically re-calculated every time a component is filtered out or added to the visualization. In Figure \ref{fig:analysis}, we can see that \texttt{krovetz} has a bigger rectangle than \texttt{lovins} and \texttt{nolug} (meaning no stemmer) showing the positive effect of the \texttt{krovetz} stemmer when interacting with the \texttt{indri} stoplist and the selected models. 

The same idea is applied to the link size: the thicker the line the better the interaction between the components it connects. For instance, in Figure \ref{fig:analysis} we can see that the stoplist-stemmer pair \texttt{indri-krovetz} has higher performances than the pair \texttt{lucene-krovetz}.

\begin{figure}[th!]
\centering
\includegraphics[width=.8\textwidth]{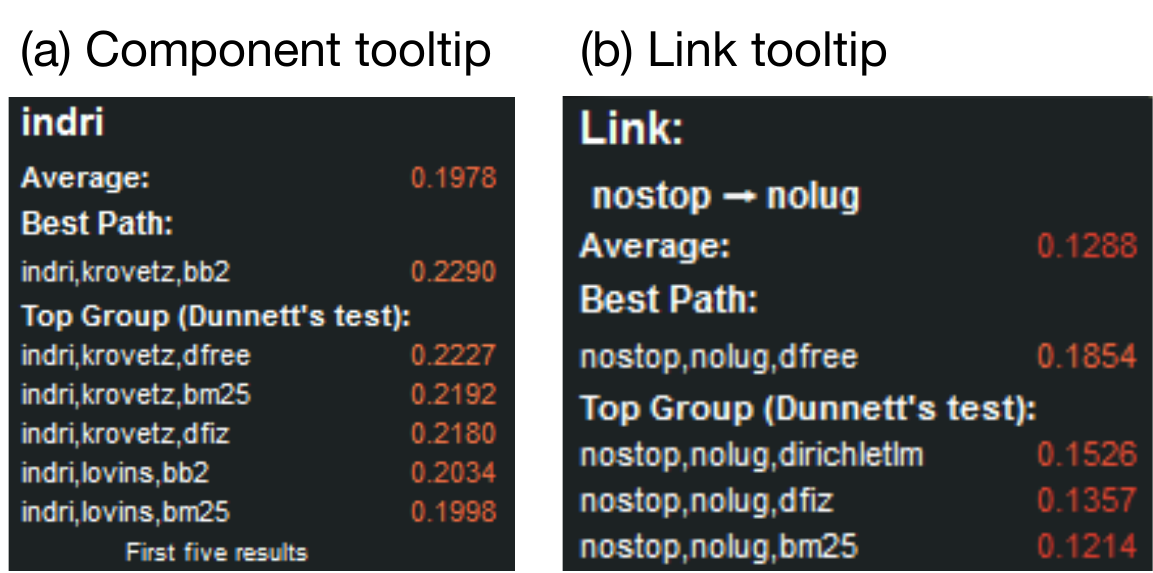}
\caption{(a) The  tooltip visualized with a mouse-over action on the \texttt{indri} stoplist component and (b) the tooltip visualized with a mouse-over the \texttt{nostop-nolug} link. \label{fig:tooltip}}

\end{figure}

With a mouse-over action on a rectangle or a link, a tooltip reporting the top 5 systems using the selected component (rectangle) or the selected components pair (link) is visualized to the user. The InfoVis system also runs the Dunnett~\cite{Dunnett1955} statistical test to determine if the reported means are statistically different one from the other. In Figure \ref{fig:tooltip}(a) we can see the tooltip visualized when the \texttt{indri} stoplist is selected: we show the average measure (AP in this case) of all the system using this stoplist, the best system adopting the stoplist and the top group of system adopting the \texttt{indri} stoplist that are not statistically different one from the other. In Figure \ref{fig:tooltip}(b) we see the tooltip reporting the statistical information related to the \texttt{nostop-nolug} link.

\section{User Evaluation}
We did a user study with nine users with a basic knowledge and previous experience with IR systems and experimental evaluation in the field; the study had a twofold goal, to compare the Sankey tool with the CLAIRE system and to conduct an in-depth analysis of the newly proposed Sankey tool. Of course, CLAIRE is a more complex system providing a wide range of functionalities, but we focused on the common features which regards the exploration of the combinatorial space of IR system pipelines.

The test was organized in three phases: (i) in-depth description of the two visual tools and hands-on phase to get to know them; (ii) \emph{comparative study:} execution of three tasks with both CLAIRE and Sankey (in this phase we divided the users into two groups where one group used firstly CLAIRE and then Sankey and the second group did the opposite); (iii) \emph{in-depth analysis:} execution of five tasks by using only the Sankey tool. 
The tasks were centered around core activities enabled by the two visual tools such as the ability to determine the best IR system, the best combination of components, the comparison between two or more alternative components and so on. 
After the resolution of the first group of tasks the users were required to fill closed questionnaire.  After the resolution of the second group of tasks the users were required to fill in an open questionnaire.

The questionnaire relative to the first set of tasks required to get a preference between Sankey and CLAIRE, was composed of two sets of questions; the first set with three questions: (Q1) How intuitive was the Sankey (CLAIRE) tool? (Q2) In your opinion how much useful is the Sankey (CLAIRE) tool to understand the performances of IR systems? (Q3) How much effective was the Sankey (CLAIRE) tool to solve the given tasks? Each question of the questionnaire had to be answered by using an interval Likert scale ranging from 1 to 5 in which each numerical score was labeled with a description: \{1: not at all, 2: a little, 3: enough, 4: a lot, 5: quite a lot\}.
\begin{figure}[th!]
\centering
\includegraphics[width=.8\textwidth]{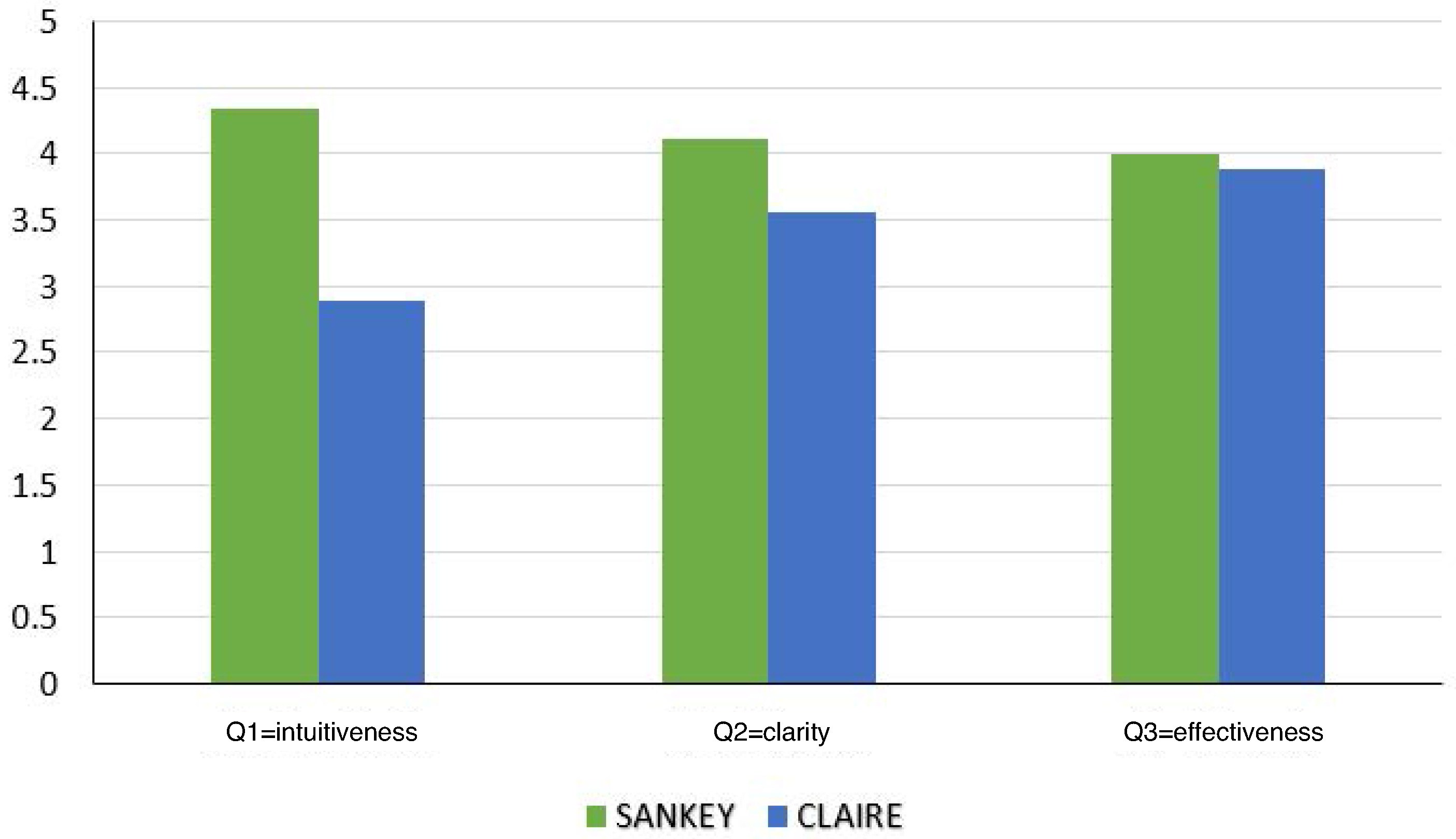}
\caption{Average answers for the first set of questions of the comparative study \label{fig:comparative}}
\end{figure}

In Figure \ref{fig:comparative} we can see that both systems were evaluated as clear to use (Q2) and effective (Q3) in both cases with a slight preference for Sankey; but Sankey was considered more intuitive than CLAIRE (Q1).

The second set of questions for the comparative study was: (Q1) Which system does represent better the experimental data? (Q2) Which system does offer the most intuitive interface to interact with the data? (Q3) Which system is more complete to solve the assigned tasks? (Q4) Which system did you prefer to use? Each question of the questionnaires had to be answered by indicating a strong preference (a ``2'' in our interval scale) or a mild preference (a ``1'') for CLAIRE or Sankey where a ``0'' value indicated equality between the systems.
 
\begin{figure}[th!]
\centering
\includegraphics[width=.45\textwidth]{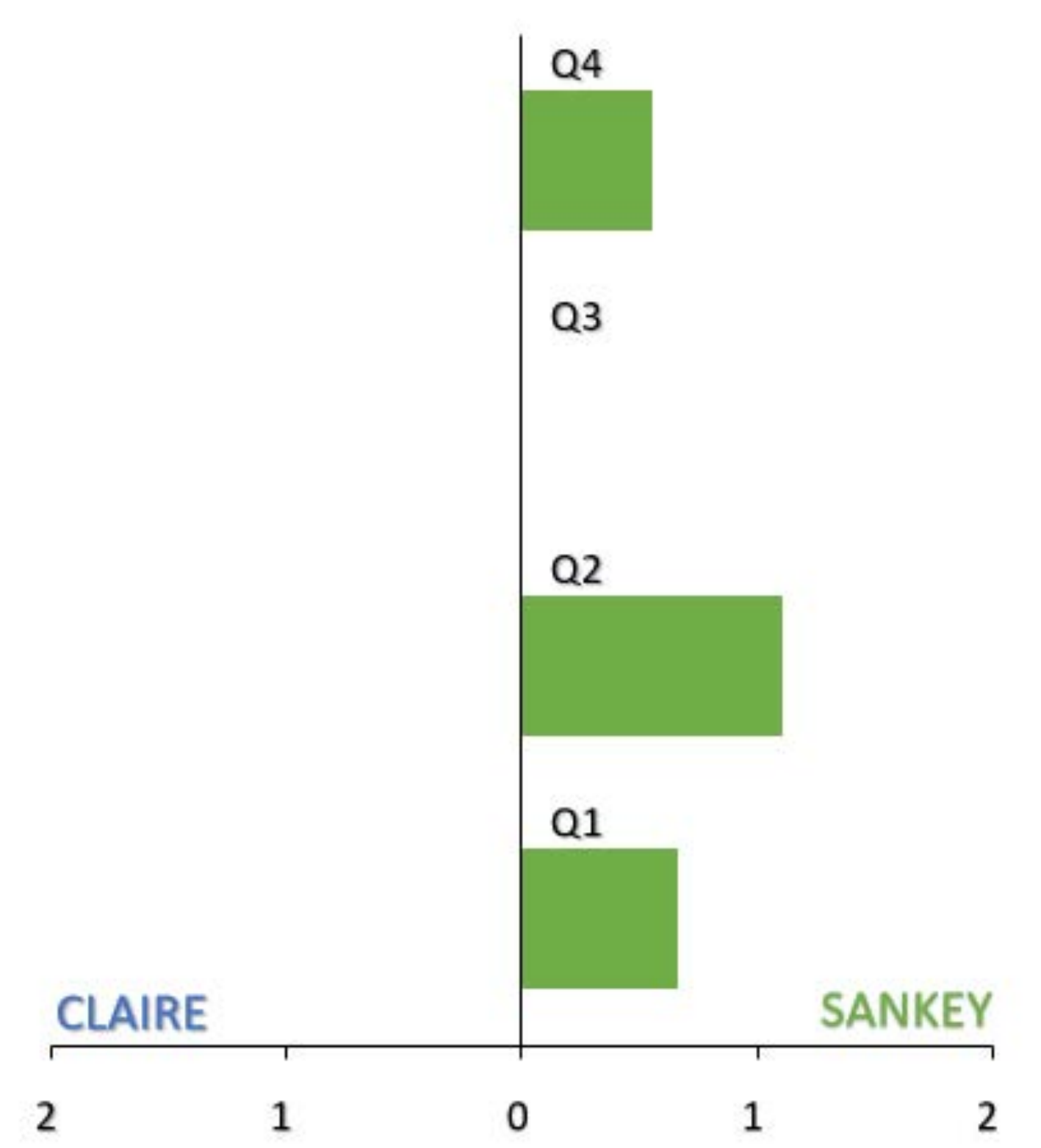}
\caption{Average answers for the second set of questions of the comparative study \label{fig:comparative2}}

\end{figure}

In Figure \ref{fig:comparative2} we can see that on average Sankey was preferred by the users with the only exception of Q3 where the systems were judged equivalent.

\section{Final remarks}
The InfoVis tool we presented has the goal to ease the exploration and analysis of large experimental \ac{GoP} enabling \ac{IR} researchers and practitioners to better understand the performances of single components, their interactions and their impact on off-the-shelf IR systems. The InfoVis tool we propose is highly interactive and remarkably simple as shown by the user study we conducted, yet offering advanced statistical information and analytics functionalities. 

The presented tool is available on-line at the URL \url{http://gridofpoints.dei.unipd.it/sankey/} and the source code is openly shared at the URL \url{https://github.com/giansilv/sankey_eval}.

\acrodef{3G}[3G]{Third Generation Mobile System}
\acrodef{5S}[5S]{Streams, Structures, Spaces, Scenarios, Societies}
\acrodef{AAAI}[AAAI]{Association for the Advancement of Artificial Intelligence}
\acrodef{AAL}[AAL]{Annotation Abstraction Layer}
\acrodef{AAM}[AAM]{Automatic Annotation Manager}
\acrodef{ACLIA}[ACLIA]{Advanced Cross-Lingual Information Access}
\acrodef{ACM}[ACM]{Association for Computing Machinery}
\acrodef{ADSL}[ADSL]{Asymmetric Digital Subscriber Line}
\acrodef{ADUI}[ADUI]{ADministrator User Interface}
\acrodef{AIP}[AIP]{Archival Information Package}
\acrodef{AJAX}[AJAX]{Asynchronous JavaScript Technology and \acs{XML}}
\acrodef{ALU}[ALU]{Aritmetic-Logic Unit}
\acrodef{AMUSID}[AMUSID]{Adaptive MUSeological IDentity-service}
\acrodef{ANOVA}[ANOVA]{ANalysis Of VAriance}
\acrodef{ANSI}[ANSI]{American National Standards Institute}
\acrodef{AP}[AP]{Average Precision}
\acrodef{APC}[APC]{AP Correlation}
\acrodef{API}[API]{Application Program Interface}
\acrodef{AR}[AR]{Address Register}
\acrodef{AS}[AS]{Annotation Service}
\acrodef{ASAP}[ASAP]{Adaptable Software Architecture Performance}
\acrodef{ASI}[ASI]{Annotation Service Integrator}
\acrodef{ASM}[ASM]{Annotation Storing Manager}
\acrodef{ASR}[ASR]{Automatic Speech Recognition}
\acrodef{ASUI}[ASUI]{ASsessor User Interface}
\acrodef{ATIM}[ATIM]{Annotation Textual Indexing Manager}
\acrodef{AUC}[AUC]{Area Under the ROC Curve}
\acrodef{AUI}[AUI]{Administrative User Interface}
\acrodef{AWARE}[AWARE]{Assessor-driven Weighted Averages for Retrieval Evaluation}
\acrodef{BANKS-I}[BANKS-I]{Browsing ANd Keyword Searching I}
\acrodef{BANKS-II}[BANKS-II]{Browsing ANd Keyword Searching II}
\acrodef{bpref}[bpref]{Binary Preference}
\acrodef{BNF}[BNF]{Backus and Naur Form}
\acrodef{BRICKS}[BRICKS]{Building Resources for Integrated Cultural Knowledge Services}
\acrodef{CAN}[CAN]{Content Addressable Netword}
\acrodef{CAS}[CAS]{Content-And-Structure}
\acrodef{CBSD}[CBSD]{Component-Based Software Developlement}
\acrodef{CBSE}[CBSE]{Component-Based Software Engineering}
\acrodef{CB-SPE}[CB-SPE]{Component-Based \acs{SPE}}
\acrodef{CD}[CD]{Collaboration Diagram}
\acrodef{CD}[CD]{Compact Disk}
\acrodef{CENL}[CENL]{Conference of European National Librarians}
\acrodef{CIDOC CRM}[CIDOC CRM]{CIDOC Conceptual Reference Model}
\acrodef{CIR}[CIR]{Current Instruction Register}
\acrodef{CIRCO}[CIRCO]{Coordinated Information Retrieval Components Orchestration}
\acrodef{CG}[CG]{Cumulated Gain}
\acrodef{CLAIRE}[CLAIRE]{Combinatorial visuaL Analytics system for Information Retrieval Evaluation}
\acrodef{CLEF}[CLEF]{Conference and Labs of the Evaluation Forum}
\acrodef{CLIR}[CLIR]{Cross Language Information Retrieval}
\acrodef{CMS}[CMS]{Content Management System}
\acrodef{CMT}[CMT]{Campaign Management Tool}
\acrodef{CNR}[CNR]{Italian National Council of Research}
\acrodef{CO}[CO]{Content-Only}
\acrodef{COD}[COD]{Code On Demand}
\acrodef{CODATA}[CODATA]{Committee on Data for Science and Technology}
\acrodef{COLLATE}[COLLATE]{Collaboratory for Annotation Indexing and Retrieval of Digitized Historical Archive Material}
\acrodef{CP}[CP]{Characteristic Pattern}
\acrodef{CPE}[CPE]{Control Processor Element}
\acrodef{CPU}[CPU]{Central Processing Unit}
\acrodef{CQL}[CQL]{Contextual Query Language}
\acrodef{CRP}[CRP]{Cumulated Relative Position}
\acrodef{CRUD}[CRUD]{Create--Read--Update--Delete}
\acrodef{CS}[CS]{Characteristic Structure}
\acrodef{CSM}[CSM]{Campaign Storing Manager}
\acrodef{CSS}[CSS]{Cascading Style Sheets}
\acrodef{CU}[CU]{Control Unit}
\acrodef{CUI}[CUI]{Client User Interface}
\acrodef{CV}[CV]{Cross-Validation}
\acrodef{DAFFODIL}[DAFFODIL]{Distributed Agents for User-Friendly Access of Digital Libraries}
\acrodef{DAO}[DAO]{Data Access Object}
\acrodef{DARE}[DARE]{Drawing Adequate REpresentations}
\acrodef{DARPA}[DARPA]{Defense Advanced Research Projects Agency}
\acrodef{DAS}[DAS]{Distributed Annotation System}
\acrodef{DB}[DB]{DataBase}
\acrodef{DBMS}[DBMS]{DataBase Management System}
\acrodef{DC}[DC]{Dublin Core}
\acrodef{DCG}[DCG]{Discounted Cumulated Gain}
\acrodef{DCMI}[DCMI]{Dublin Core Metadata Initiative}
\acrodef{DCV}[DCV]{Document Cut--off Value}
\acrodef{DD}[DD]{Deployment Diagram}
\acrodef{DDC}[DDC]{Dewey Decimal Classification}
\acrodef{DDS}[DDS]{Direct Data Structure}
\acrodef{DF}[DF]{Degrees of Freedom}
\acrodef{DFR}[DFR]{Divergence From Randomness}
\acrodef{DHT}[DHT]{Distributed Hash Table}
\acrodef{DI}[DI]{Digital Image}
\acrodef{DIKW}[DIKW]{Data, Information, Knowledge, Wisdom}
\acrodef{DIL}[DIL]{\acs{DIRECT} Integration Layer}
\acrodef{DiLAS}[DiLAS]{Digital Library Annotation Service}
\acrodef{DIRECT}[DIRECT]{Distributed Information Retrieval Evaluation Campaign Tool}
\acrodef{DKMS}[DKMS]{Data and Knowledge Management System}
\acrodef{DL}[DL]{Digital Library}
\acrodefplural{DL}[DL]{Digital Libraries}
\acrodef{DLMS}[DLMS]{Digital Library Management System}
\acrodef{DLOG}[DL]{Description Logics}
\acrodef{DLS}[DLS]{Digital Library System}
\acrodef{DLSS}[DLSS]{Digital Library Service System}
\acrodef{DO}[DO]{Digital Object}
\acrodef{DOI}[DOI]{Digital Object Identifier}
\acrodef{DOM}[DOM]{Document Object Model}
\acrodef{DoMDL}[DoMDL]{Document Model for Digital Libraries}
\acrodef{DPBF}[DPBF]{Dynamic Programming Best-First}
\acrodef{DR}[DR]{Data Register}
\acrodef{DRIVER}[DRIVER]{Digital Repository Infrastructure Vision for European Research}
\acrodef{DTD}[DTD]{Document Type Definition}
\acrodef{DVD}[DVD]{Digital Versatile Disk}
\acrodef{EAC-CPF}[EAC-CPF]{Encoded Archival Context for Corporate Bodies, Persons, and Families}
\acrodef{EAD}[EAD]{Encoded Archival Description}
\acrodef{EAN}[EAN]{International Article Number}
\acrodef{ECD}[ECD]{Enhanced Contenty Delivery}
\acrodef{ECDL}[ECDL]{European Conference on Research and Advanced Technology for Digital Libraries}
\acrodef{EDM}[EDM]{Europeana Data Model}
\acrodef{EG}[EG]{Execution Graph}
\acrodef{ELDA}[ELDA]{Evaluation and Language resources Distribution Agency}
\acrodef{ELRA}[ELRA]{European Language Resources Association}
\acrodef{EM}[EM]{Expectation Maximization}
\acrodef{EMMA}[EMMA]{Extensible MultiModal Annotation}
\acrodef{EPROM}[EPROM]{Erasable Programmable \acs{ROM}}
\acrodef{EQNM}[EQNM]{Extended Queueing Network Model}
\acrodef{ER}[ER]{Entity--Relationship}
\acrodef{ERR}[ERR]{Expected Reciprocal Rank}
\acrodef{ETL}[ETL]{Extract-Transform-Load}
\acrodef{FAST}[FAST]{Flexible Annotation Service Tool}
\acrodef{FIFO}[FIFO]{First-In / First-Out}
\acrodef{FIRE}[FIRE]{Forum for Information Retrieval Evaluation}
\acrodef{FN}[FN]{False Negative}
\acrodef{FNR}[FNR]{False Negative Rate}
\acrodef{FOAF}[FOAF]{Friend of a Friend}
\acrodef{FORESEE}[FORESEE]{FOod REcommentation sErvER}
\acrodef{FP}[FP]{False Positive}
\acrodef{FPR}[FPR]{False Positive Rate}
\acrodef{GIF}[GIF]{Graphics Interchange Format}
\acrodef{GIR}[GIR]{Geografic Information Retrieval}
\acrodef{GAP}[GAP]{Graded Average Precision}
\acrodef{GLM}[GLM]{General Linear Model}
\acrodef{GLMM}[GLMM]{General Linear Mixed Model}
\acrodef{GMAP}[GMAP]{Geometric Mean Average Precision}
\acrodef{GoP}[GoP]{Grid of Points}
\acrodef{GPRS}[GPRS]{General Packet Radio Service}
\acrodef{gRBP}[gRBP]{Graded Rank-Biased Precision}
\acrodef{GTIN}[GTIN]{Global Trade Item Number}
\acrodef{GUI}[GUI]{Graphical User Interface}
\acrodef{GW}[GW]{Gateway}
\acrodef{HCI}[HCI]{Human Computer Interaction}
\acrodef{HDS}[HDS]{Hybrid Data Structure}
\acrodef{HIR}[HIR]{Hypertext Information Retrieval}
\acrodef{HIT}[HIT]{Human Intelligent Task}
\acrodef{HITS}[HITS]{Hyperlink-Induced Topic Search}
\acrodef{HTML}[HTML]{HyperText Markup Language}
\acrodef{HTTP}[HTTP]{HyperText Transfer Protocol}
\acrodef{HSD}[HSD]{Honestly Significant Difference}
\acrodef{ICA}[ICA]{International Council on Archives}
\acrodef{ICSU}[ICSU]{International Council for Science}
\acrodef{IDF}[IDF]{Inverse Document Frequency}
\acrodef{IDS}[IDS]{Inverse Data Structure}
\acrodef{IEEE}[IEEE]{Institute of Electrical and Electronics Engineers}
\acrodef{IEI}[IEI]{Istituto della Enciclopedia Italiana fondata da Giovanni Treccani}
\acrodef{IETF}[IETF]{Internet Engineering Task Force}
\acrodef{IMS}[IMS]{Information Management System}
\acrodef{IMSPD}[IMS]{Information Management Systems Research Group}
\acrodef{indAP}[indAP]{Induced Average Precision}
\acrodef{infAP}[infAP]{Inferred Average Precision}
\acrodef{INEX}[INEX]{INitiative for the Evaluation of \acs{XML} Retrieval}
\acrodef{INS-M}[INS-M]{Inverse Set Data Model}
\acrodef{INTR}[INTR]{Interrupt Register}
\acrodef{IP}[IP]{Internet Protocol}
\acrodef{IPSA}[IPSA]{Imaginum Patavinae Scientiae Archivum}
\acrodef{IR}[IR]{Information Retrieval}
\acrodef{IRON}[IRON]{Information Retrieval ON}
\acrodef{IRON2}[IRON$^2$]{Information Retrieval On aNNotations}
\acrodef{IRON-SAT}[IRON-SAT]{\acs{IRON} - Statistical Analysis Tool}
\acrodef{IRS}[IRS]{Information Retrieval System}
\acrodef{ISAD(G)}[ISAD(G)]{International Standard for Archival Description (General)}
\acrodef{ISBN}[ISBN]{International Standard Book Number}
\acrodef{ISIS}[ISIS]{Interactive SImilarity Search}
\acrodef{ISJ}[ISJ]{Interactive Searching and Judging}
\acrodef{ISO}[ISO]{International Organization for Standardization}
\acrodef{ITU}[ITU]{International Telecommunication Union }
\acrodef{ITU-T}[ITU-T]{Telecommunication Standardization Sector of \acs{ITU}}
\acrodef{IV}[IV]{Information Visualization}
\acrodef{JAN}[JAN]{Japanese Article Number}
\acrodef{JDBC}[JDBC]{Java DataBase Connectivity}
\acrodef{JMB}[JMB]{Java--Matlab Bridge}
\acrodef{JPEG}[JPEG]{Joint Photographic Experts Group}
\acrodef{JSON}[JSON]{JavaScript Object Notation}
\acrodef{JSP}[JSP]{Java Server Pages}
\acrodef{JTE}[JTE]{Java-Treceval Engine}
\acrodef{KDE}[KDE]{Kernel Density Estimation}
\acrodef{KLD}[KLD]{Kullback-Leibler Divergence}
\acrodef{KLAPER}[KLAPER]{Kernel LAnguage for PErformance and Reliability analysis}
\acrodef{LAM}[LAM]{Libraries, Archives, and Museums}
\acrodef{LAM2}[LAM]{Logistic Average Misclassification}
\acrodef{LAN}[LAN]{Local Area Network}
\acrodef{LD}[LD]{Linked Data}
\acrodef{LEAF}[LEAF]{Linking and Exploring Authority Files}
\acrodef{LIDO}[LIDO]{Lightweight Information Describing Objects}
\acrodef{LIFO}[LIFO]{Last-In / First-Out}
\acrodef{LM}[LM]{Language Model}
\acrodef{LMT}[LMT]{Log Management Tool}
\acrodef{LOD}[LOD]{Linked Open Data}
\acrodef{LODE}[LODE]{Linking Open Descriptions of Events}
\acrodef{LpO}[LpO]{Leave-$p$-Out}
\acrodef{LRM}[LRM]{Local Relational Model}
\acrodef{LRU}[LRU]{Last Recently Used}
\acrodef{LS}[LS]{Lexical Signature}
\acrodef{LSM}[LSM]{Log Storing Manager}
\acrodef{LUG}[LUG]{Lexical Unit Generator}
\acrodef{MA}[MA]{Mobile Agent}
\acrodef{MA}[MA]{Moving Average}
\acrodef{MACS}[MACS]{Multilingual ACcess to Subjects}
\acrodef{MADCOW}[MADCOW]{Multimedia Annotation of Digital Content Over the Web}
\acrodef{MAD}[MAD]{Mean Assessed Documents}
\acrodef{MADP}[MADP]{Mean Assessed Documents Precision}
\acrodef{MADS}[MADS]{Metadata Authority Description Standard}
\acrodef{MAP}[MAP]{Mean Average Precision}
\acrodef{MARC}[MARC]{Machine Readable Cataloging}
\acrodef{MATTERS}[MATTERS]{MATlab Toolkit for Evaluation of information Retrieval Systems}
\acrodef{MDA}[MDA]{Model Driven Architecture}
\acrodef{MDD}[MDD]{Model-Driven Development}
\acrodef{METS}[METS]{Metadata Encoding and Transmission Standard}
\acrodef{MIDI}[MIDI]{Musical Instrument Digital Interface}
\acrodef{MIME}[MIME]{Multipurpose Internet Mail Extensions}
\acrodef{MLIA}[MLIA]{MultiLingual Information Access}
\acrodef{MM}[MM]{Machinery Model}
\acrodef{MMU}[MMU]{Memory Management Unit}
\acrodef{MODS}[MODS]{Metadata Object Description Schema}
\acrodef{MOF}[MOF]{Meta-Object Facility}
\acrodef{MP}[MP]{Markov Precision}
\acrodef{MPEG}[MPEG]{Motion Picture Experts Group}
\acrodef{MRD}[MRD]{Machine Readable Dictionary}
\acrodef{MRF}[MRF]{Markov Random Field}
\acrodef{MS}[MS]{Mean Squares}
\acrodef{MSAC}[MSAC]{Multilingual Subject Access to Catalogues}
\acrodef{MSE}[MSE]{Mean Square Error}
\acrodef{MT}[MT]{Machine Translation}
\acrodef{MV}[MV]{Majority Vote}
\acrodef{MVC}[MVC]{Model-View-Controller}
\acrodef{NACSIS}[NACSIS]{NAtional Center for Science Information Systems}
\acrodef{NAP}[NAP]{Network processors Applications Profile}
\acrodef{NCP}[NCP]{Normalized Cumulative Precision}
\acrodef{nCG}[nCG]{Normalized Cumulated Gain}
\acrodef{nCRP}[nCRP]{Normalized Cumulated Relative Position}
\acrodef{nDCG}[nDCG]{Normalized Discounted Cumulated Gain}
\acrodef{NESTOR}[NESTOR]{NEsted SeTs for Object hieRarchies}
\acrodef{NEXI}[NEXI]{Narrowed Extended XPath I}
\acrodef{NII}[NII]{National Institute of Informatics}
\acrodef{NISO}[NISO]{National Information Standards Organization}
\acrodef{NIST}[NIST]{National Institute of Standards and Technology}
\acrodef{NLP}[NLP]{Natural Language Processing}
\acrodef{NP}[NP]{Network Processor}
\acrodef{NR}[NR]{Normalized Recall}
\acrodef{NS-M}[NS-M]{Nested Set Model}
\acrodef{NTCIR}[NTCIR]{NII Testbeds and Community for Information access Research}
\acrodef{OAI}[OAI]{Open Archives Initiative}
\acrodef{OAI-ORE}[OAI-ORE]{Open Archives Initiative Object Reuse and Exchange}
\acrodef{OAI-PMH}[OAI-PMH]{Open Archives Initiative Protocol for Metadata Harvesting}
\acrodef{OAIS}[OAIS]{Open Archival Information System}
\acrodef{OC}[OC]{Operation Code}
\acrodef{OCLC}[OCLC]{Online Computer Library Center}
\acrodef{OMG}[OMG]{Object Management Group}
\acrodef{OLAP}[OLAP]{On-Line Analytical Processing}
\acrodef{OO}[OO]{Object Oriented}
\acrodef{OODB}[OODB]{Object-Oriented \acs{DB}}
\acrodef{OODBMS}[OODBMS]{Object-Oriented \acs{DBMS}}
\acrodef{OPAC}[OPAC]{Online Public Access Catalog}
\acrodef{OQL}[OQL]{Object Query Language}
\acrodef{ORP}[ORP]{Open Relevance Project}
\acrodef{OSIRIS}[OSIRIS]{Open Service Infrastructure for Reliable and Integrated process Support}
\acrodef{P2P}[P2P]{Peer-To-Peer}
\acrodef{PA}[PA]{Performance Analysis}
\acrodef{PAMT}[PAMT]{Pool-Assessment Management Tool}
\acrodef{PASM}[PASM]{Pool-Assessment Storing Manager}
\acrodef{PC}[PC]{Program Counter}
\acrodef{PCP}[PCP]{Pre-Commercial Procurement}
\acrodef{PCR}[PCR]{Peripherical Command Register}
\acrodef{PDA}[PDA]{Personal Digital Assistant}
\acrodef{PDF}[PDF]{Probability Density Function}
\acrodef{PDR}[PDR]{Peripherical Data Register}
\acrodef{POI}[POI]{\acs{PURL}-based Object Identifier}
\acrodef{PoS}[PoS]{Part of Speech}
\acrodef{PPE}[PPE]{Programmable Processing Engine}
\acrodef{PREFORMA}[PREFORMA]{PREservation FORMAts for culture information/e-archives}
\acrodef{PRIMAmob-UML}[PRIMAmob-UML]{mobile \acs{PRIMA-UML}}
\acrodef{PRIMA-UML}[PRIMA-UML]{PeRformance IncreMental vAlidation in \acs{UML}}
\acrodef{PROM}[PROM]{Programmable \acs{ROM}}
\acrodef{PROMISE}[PROMISE]{Participative Research labOratory  for Multimedia and Multilingual Information Systems Evaluation}
\acrodef{pSQL}[pSQL]{propagate \acs{SQL}}
\acrodef{PUI}[PUI]{Participant User Interface}
\acrodef{PURL}[PURL]{Persistent \acs{URL}}
\acrodef{QA}[QA]{Question Answering}
\acrodef{QoS-UML}[QoS-UML]{\acs{UML} Profile for QoS and Fault Tolerance}
\acrodef{RAM}[RAM]{Random Access Memory}
\acrodef{RAMM}[RAM]{Random Access Machine}
\acrodef{RBO}[RBO]{Rank-Biased Overlap}
\acrodef{RBP}[RBP]{Rank-Biased Precision}
\acrodef{RDBMS}[RDBMS]{Relational \acs{DBMS}}
\acrodef{RDF}[RDF]{Resource Description Framework}
\acrodef{REST}[REST]{REpresentational State Transfer}
\acrodef{REV}[REV]{Remote Evaluation}
\acrodef{RFC}[RFC]{Request for Comments}
\acrodef{RIA}[RIA]{Reliable Information Access}
\acrodef{RMSE}[RMSE]{Root Mean Square Error}
\acrodef{RMT}[RMT]{Run Management Tool}
\acrodef{ROM}[ROM]{Read Only Memory}
\acrodef{ROMIP}[ROMIP]{Russian Information Retrieval Evaluation Seminar}
\acrodef{RoMP}[RoMP]{Rankings of Measure Pairs}
\acrodef{RoS}[RoS]{Rankings of Systems}
\acrodef{RP}[RP]{Relative Position}
\acrodef{RR}[RR]{Reciprocal Rank}
\acrodef{RSM}[RSM]{Run Storing Manager}
\acrodef{RST}[RST]{Rhetorical Structure Theory}
\acrodef{RT-UML}[RT-UML]{\acs{UML} Profile for Schedulability, Performance and Time}
\acrodef{SA}[SA]{Software Architecture}
\acrodef{SAL}[SAL]{Storing Abstraction Layer}
\acrodef{SAMT}[SAMT]{Statistical Analysis Management Tool}
\acrodef{SAN}[SAN]{Sistema Archivistico Nazionale}
\acrodef{SASM}[SASM]{Statistical Analysis Storing Manager}
\acrodef{SD}[SD]{Sequence Diagram}
\acrodef{SE}[SE]{Search Engine}
\acrodef{SEBD}[SEBD]{Convegno Nazionale su Sistemi Evoluti per Basi di Dati}
\acrodef{SFT}[SFT]{Satisfaction--Frustration--Total}
\acrodef{SIL}[SIL]{Service Integration Layer}
\acrodef{SIP}[SIP]{Submission Information Package}
\acrodef{SKOS}[SKOS]{Simple Knowledge Organization System}
\acrodef{SM}[SM]{Software Model}
\acrodef{SMART}[SMART]{System for the Mechanical Analysis and Retrieval of Text}
\acrodef{SoA}[SoA]{Service-oriented Architectures}
\acrodef{SOA}[SOA]{Strength of Association}
\acrodef{SOAP}[SOAP]{Simple Object Access Protocol}
\acrodef{SOM}[SOM]{Self-Organizing Map}
\acrodef{SPE}[SPE]{Software Performance Engineering}
\acrodef{SPINA}[SPINA]{Superimposed Peer Infrastructure for iNformation Access}
\acrodef{SPLIT}[SPLIT]{Stemming Program for Language Independent Tasks}
\acrodef{SPOOL}[SPOOL]{Simultaneous Peripheral Operations On Line}
\acrodef{SQL}[SQL]{Structured Query Language}
\acrodef{SR}[SR]{Sliding Ratio}
\acrodef{SR}[SR]{Status Register}
\acrodef{SRU}[SRU]{Search/Retrieve via \acs{URL}}
\acrodef{SS}[SS]{Sum of Squares}
\acrodef{SSTF}[SSTF]{Shortest Seek Time First}
\acrodef{STAR}[STAR]{Steiner-Tree Approximation in Relationship graphs}
\acrodef{STON}[STON]{STemming ON}
\acrodef{TAC}[TAC]{Text Analysis Conference}
\acrodef{TBG}[TBG]{Time-Biased Gain}
\acrodef{TCP}[TCP]{Transmission Control Protocol}
\acrodef{TEL}[TEL]{The European Library}
\acrodef{TERRIER}[TERRIER]{TERabyte RetrIEveR}
\acrodef{TF}[TF]{Term Frequency}
\acrodef{TFR}[TFR]{True False Rate}
\acrodef{TN}[TN]{True Negative}
\acrodef{TO}[TO]{Transfer Object}
\acrodef{TP}[TP]{True Positve}
\acrodef{TPR}[TPR]{True Positive Rate}
\acrodef{TRAT}[TRAT]{Text Relevance Assessing Task}
\acrodef{TREC}[TREC]{Text REtrieval Conference}
\acrodef{TRECVID}[TRECVID]{TREC Video Retrieval Evaluation}
\acrodef{TTL}[TTL]{Time-To-Live}
\acrodef{UCD}[UCD]{Use Case Diagram}
\acrodef{UDC}[UDC]{Universal Decimal Classification}
\acrodef{uGAP}[uGAP]{User-oriented Graded Average Precision}
\acrodef{UI}[UI]{User Interface}
\acrodef{UML}[UML]{Unified Modeling Language}
\acrodef{UMT}[UMT]{User Management Tool}
\acrodef{UMTS}[UMTS]{Universal Mobile Telecommunication System}
\acrodef{UoM}[UoM]{Utility-oriented Measurement}
\acrodef{UPC}[UPC]{Universal Product Code}
\acrodef{URI}[URI]{Uniform Resource Identifier}
\acrodef{URL}[URL]{Uniform Resource Locator}
\acrodef{URN}[URN]{Uniform Resource Name}
\acrodef{USM}[USM]{User Storing Manager}
\acrodef{VA}[VA]{Visual Analytics}
\acrodef{VATE}[VATE$^2$]{Visual Analytics Tool for Experimental Evaluation}
\acrodef{VIRTUE}[VIRTUE]{Visual Information Retrieval Tool for Upfront Evaluation}
\acrodef{VD}[VD]{Virtual Document}
\acrodef{VIAF}[VIAF]{Virtual International Authority File}
\acrodef{VL}[VL]{Visual Language}
\acrodef{VoIP}[VoIP]{Voice over IP}
\acrodef{VS}[VS]{Visual Sentence}
\acrodef{W3C}[W3C]{World Wide Web Consortium}
\acrodef{WAN}[WAN]{Wide Area Network}
\acrodef{WHO}[WHO]{World Health Organization}
\acrodef{WLAN}[WLAN]{Wireless \acs{LAN}}
\acrodef{WP}[WP]{Work Package}
\acrodef{WS}[WS]{Web Services}
\acrodef{WSD}[WSD]{Word Sense Disambiguation}
\acrodef{WSDL}[WSDL]{Web Services Description Language}
\acrodef{WWW}[WWW]{World Wide Web}
\acrodef{XMI}[XMI]{\acs{XML} Metadata Interchange}
\acrodef{XML}[XML]{eXtensible Markup Language}
\acrodef{XPath}[XPath]{XML Path Language}
\acrodef{XSL}[XSL]{eXtensible Stylesheet Language}
\acrodef{XSL-FO}[XSL-FO]{\acs{XSL} Formatting Objects}
\acrodef{XSLT}[XSLT]{\acs{XSL} Transformations}
\acrodef{YAGO}[YAGO]{Yet Another Great Ontology}
\acrodef{YASS}[YASS]{Yet Another Suffix Stripper}

\clearpage
\newpage

\end{document}